\documentclass[doublespacing]{elsart}
\usepackage{stmaryrd}
\usepackage{amsmath}
\usepackage{amssymb}
\usepackage{amsfonts}
\usepackage{graphicx}
\usepackage{subfigure}
\usepackage{rotating}
\usepackage{longtable}
\usepackage{graphics}
\usepackage{epstopdf}
\usepackage{epsfig}

\date{}
\begin{document}

\noindent \textbf{Corresponding author: }\\
\noindent Prof. Dr. Hong-Jian Feng\\
\noindent
Department of Physics,\\
Northwest University,\\ Xi'an 710069, People's Republic of China\\
Tel.:
+86-29-88303384\\
Email address:\\
hjfeng@nwu.edu.cn\\
fenghongjian@gmail.com\\
\clearpage

\begin{frontmatter}

\title{Spin transfer in ultrathin   BiFeO$_3$ film under external electric field}





\author{Hong-Jian  Feng}

\address{ Department of  Physics, Northwest University, Xi'an 710069, People's Republic of China}

\begin{abstract}

First-principals calculations show that up-spin and down-spin carriers are accumulating   adjacent to opposite surfaces of  BiFeO$_3$(BFO) film   with applying external bias. The spin carriers are equal in magnitude and opposite in direction, and down-spin carriers move to the direction opposing the external electric field while up-spin ones along the field direction. This novel spin transfer properties make BFO film an intriguing candidate for application in spin capacitor and BFO-based multiferroic field-effect device.
\end{abstract}

\end{frontmatter}






 BiFeO$_3$(BFO) has attracted much interests due to the coupling between antiferromagnetic and ferroelectric order parameters at room-temperature which drives  a potential application in magnetic memory storage and emerging spintronics\cite{1,2}. The electromechanical coupling across a morphotropic phase boundary makes it a probe-based data storage candidate\cite{3}. In addition the diode effect, as well as the visible light photovoltaic effect, observed in BFO open a new way toward novel optoelectronics applications\cite{4}. An electric-field controllable magnetization can be realized by producing a heterointerface composed by the BFO antiferromagnetic layer and ferromagnetic layer,such as BFO-La$_{0.7}$Sr$_{0.3}$MnO(LSMO) heterostructure\cite{5,6}, through exchange-bias effect in the interface, instead of using the single-phase multiferroics which show a weak magnetoelectric coupling effect\cite{7,8}. Surface property in BFO,specifically under external voltage, plays a  crucial role in the orbital reconstruction proposed in the BFO-LSMO heterointerface, as well as in the BFO-based multifunctional devices. Moreover, we speculate the charge transfer as well as the spin transfer also contribute to the BFO-based field emission device. Therefore the work function, as well as the spin density profile of BFO film under external electric field would give a transparent explanation for the transport properties in BFO-based heterostructures. As far as we know, associated works employing first principles calculations have not been reported in the literature, although phase transition under compressive strain\cite{9,10}, as well as the Monte Carlo\cite{11,12,13} and phase-field simulations under electric field   where Li and co-workers have revealed the nanoscale control of magnetoelectric coupling in strain engineered BFO films\cite{14} and the switching of antiferromagnetic domains by mechanical stress\cite{15}, have been carried out.

In this Letter, we construct the rhombohedral $R3c$ structure(R)(0 0 1)film in hexagonal frame of reference\cite{16},the bulk phase structure, and the tetragonal $P4mm$ structure(T)(0 0 1)film,observed in BFO  film\cite{17}.  The slab model with nine and twelve atomic layers is used to simulate the  ultrathin film for P and R phase, respectively\cite{18,19,20} and the same depth is arranged for the vacuum region. The relaxation is carried out with the top and bottom two atomic layers being moved when the forces on the ions are less than 0.01 eV/\AA. We adopt the local spin density approximation(LSDA) scheme and  G-type AFM spin configuration\cite{21,22,23} for BFO film as we have used in our previous work\cite{24,25,26}.  We still include spin-orbit corrections and noncollinear magnetism. The external electric potential\cite{27,28} has the form,
\begin{equation}
V_{ext}(\textbf{r})=4\pi m(\textbf{r}/r_m-1/2),  0<\textbf{r}<r_m
\end{equation}
where m is the surface dipole density of the slab, r$_m$ is the periodic length along the direction perpendicular to the surface. We set the external electric field varying from -2 to 2 V/${\AA}$ for R phase film and from -1 to 1 V/${\AA}$ for T phase film in that the
field more than that value cause large restoring forces and overshoot in the iteration
 process, so called "charge sloshing". For the sake of simplification, the unit for external bias used here is defined as the electric field exerted on an electron.

The work function(WF) is calculated as the difference between the averaged vacuum level($\bar{\phi}$) and the Fermi energy($E_F$) of the system. The WF for these two phases is shown  in Fig.1 as a function of the external bias. The WF of the T phase film without applying electric field possesses the highest value compared with those under external electric field. There is no significant influence on the WF with the external bias reversal, and they all exhibit a relatively lower value around 4.5 eV. On the contrary, the WF is changing dramatically under alternating external bias in R phase. The WF experience a sharp energy difference as the bias is reversed  from -1 to 1 V/${\AA}$,indicating a significant change in surface electronic properties associated with the lattice distortions in the vicinity of the surface. We suggest this behavior is related to the structural change,as well as the spin transfer properties under external electric field which will be discussed in the following section.

The electrostatic potential across the film averaged over the plane parallel to the surface with and without external electric field for R phase and T phase is illustrated in Fig.2 and Fig. 3, respectively. It can be seen clearly that the potential drop occurs in the vicinity of the film due to the external bias while the inside layers are less affected due to the screening effect. The potential distributes in a broad region and fluctuates dramatically  near the surface with external bias which is caused by the deviation of surface ions under the field. It is worth mentioning that an internal field is observed in the T phase free-standing film comparing with the R phase film without turning on the external bias. Moreover the potential distribution becomes narrower in T phase applying the electric field in contrast to  that without applying electric field. This behavior is attributed to the different charged layers in T phase while the atomic layers in R phase is neutral in nature.

We go further investigation to see the spin transfer properties and report the spin density profile for R phase and T phase film under electric field in Fig. 4 and Fig. 5, respectively. It is very interesting to see that  up-spin electrons accumulate adjacent to the right plane along the direction of electric field, while the down-spin electrons move to the left plane opposing the field direction in R phase film. Moreover the magnetization caused by the spin density in the two surfaces is equal in magnitude and opposite in direction. This partly reflects that they will compensate with each other and makes the film remains neutrally without electric field while the different spin carriers moves to the opposite direction under external electric field. Although the accumulation of different spin carriers is again found in T phase film, the magnetization adjacent to the surface caused by the spin carriers are difference in magnitude. Spin density doubles in the vicinity of the left surface comparing with that close to the right surface, implying that the different spin carriers can not be neutralized without applying electric field. This behavior is attributed partly to the charged atomic layers in the T phase film, leading to the spin polarization across the film , and this further implies that the polarization under electric field is also coupled to the spin transfer.

 The spin carriers accumulating behavior applying external voltage is illustrated clearly in Fig. 6. The amount of different spin carriers aggregating in the vicinity of the opposite surfaces are equal in R phase film while it is not the case for T phase film due to the charged internal layers, and this reflects that the spin transfer is closely related to the lattice distortions as well as the domain switching. The accumulation of up-spin and down-spin electrons adjacent to the opposite surfaces leads to the spin density and magnetization near the surface. This meaningful and fantastic spin transfer properties observed in R phase BFO film can be used in two applications:(\textrm{i})the spin capacitor which accumulates the spin carriers instead of opposite charges in traditional capacitor\cite{29}; and (\textrm{ii}) BFO-based heterointerface which could tune the magnetization of the ferromagnetic layer by the external electric field through exchange-bias effect\cite{30,31}.

In summary, $Ab$ $initio$ calculations show that the R phase BFO film exhibits a significant change in work function with reversal of external bias while this can not be observed in T phase film. It relates with the ionic structure, as well as the spin density adjacent to the surface. Atomic layers in T phase film are charged, leading to the internal potential and the unbalanced spin density near the opposite surfaces. Spin carriers accumulating behavior in R phase film presents promising possibilities for spintronic applications.

\noindent\textbf{Acknowledgment}

This work was financially supported by the National Natural Science Foundation of China(NSFC) under Grant No.11247230(H.-J. F.),
and by the Science Foundation of Northwest University(Grant No.12NW12)(H.-J. F.). We thank X. Hu and Z. Jiang for fruitful discussions.




\clearpage

\raggedright \textbf{Figure captions:}

Fig.1 Relationship between work function and  the external electric field.

Fig.2  $xy$ averaged electrostatic potential across the R phase  ultrathin BFO film with and without external electric field.

Fig.3 $xy$ averaged electrostatic potential across the T phase  ultrathin BFO film with and without external electric field.

Fig.4 Distribution of magnetization across the R phase ultrathin BFO film averaged over the plane parallel to the film under electric field of E=2 V/${\AA}$.

Fig.5 Distribution of magnetization across the T phase ultrathin BFO film averaged over the plane parallel to the film under electric field of E=1 V/${\AA}$.

Fig.6 Spin transfer mechanism applying electric field for R and T phase BFO film.

\clearpage

\begin{figure}
\centering
\includegraphics{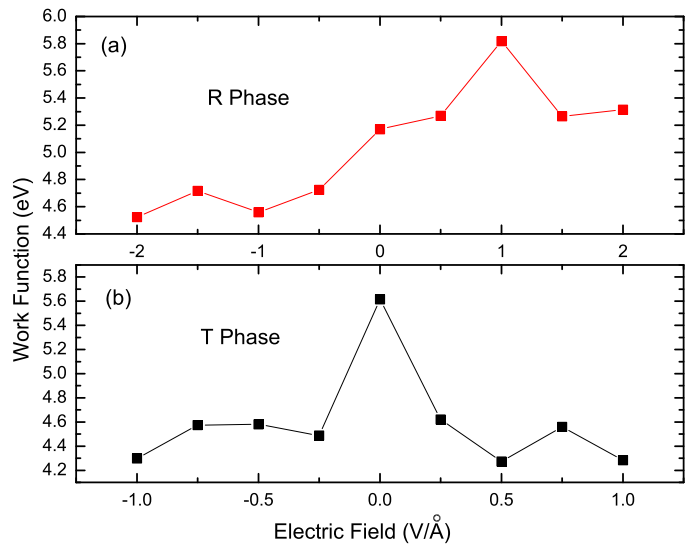}
\caption{Relationship between work function and  the external electric field.}
\end{figure}

\begin{figure}
\centering
\includegraphics{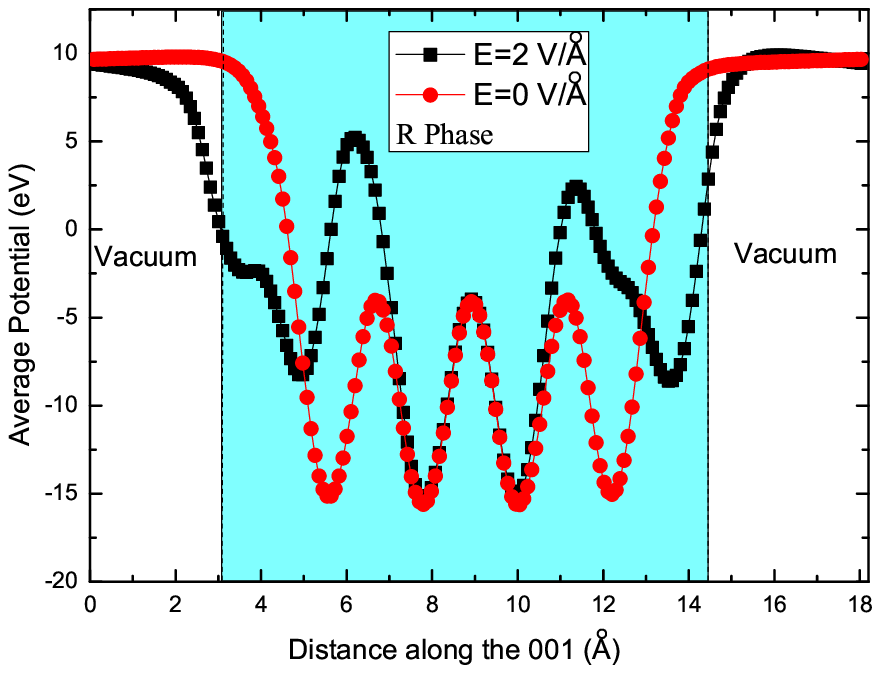}
\caption{ $xy$ averaged electrostatic potential across the R phase  ultrathin BFO film with and without external electric field.}
\end{figure}

\begin{figure}
\centering
\includegraphics{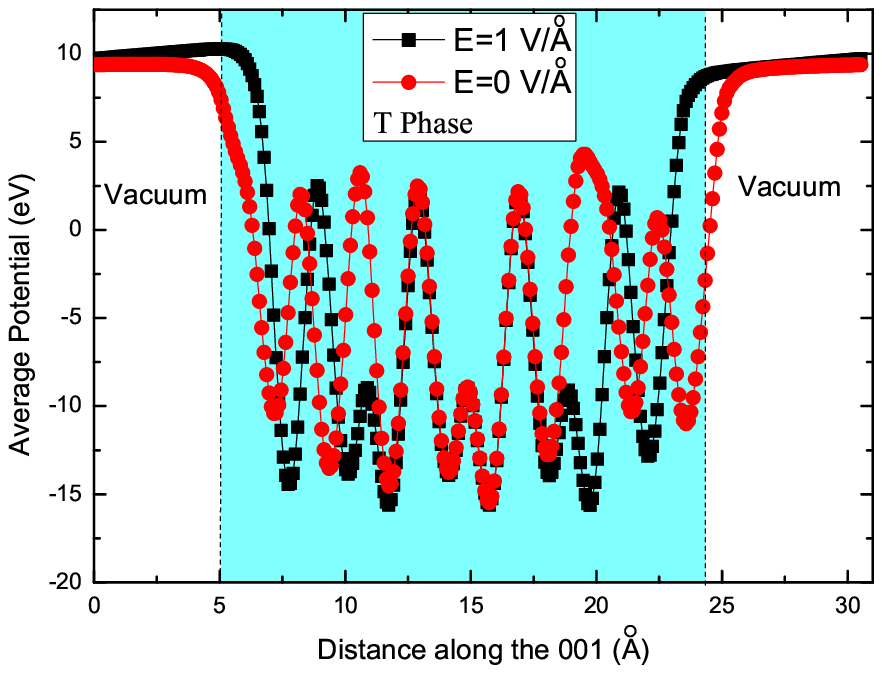}
\caption{ $xy$ averaged electrostatic potential across the T phase  ultrathin BFO film with and without external electric field.}
\end{figure}

\begin{figure}
\centering
\includegraphics{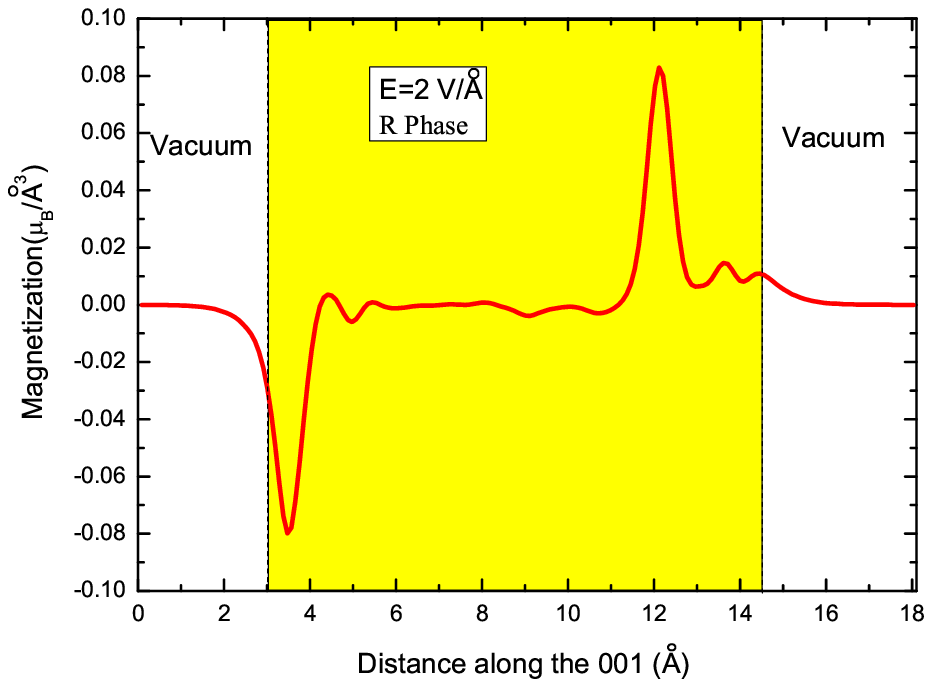}
\caption{Distribution of magnetization across the R phase ultrathin BFO film averaged over the plane parallel to the film under electric field of E=2 V/${\AA}$.}
\end{figure}

\begin{figure}
\centering
\includegraphics{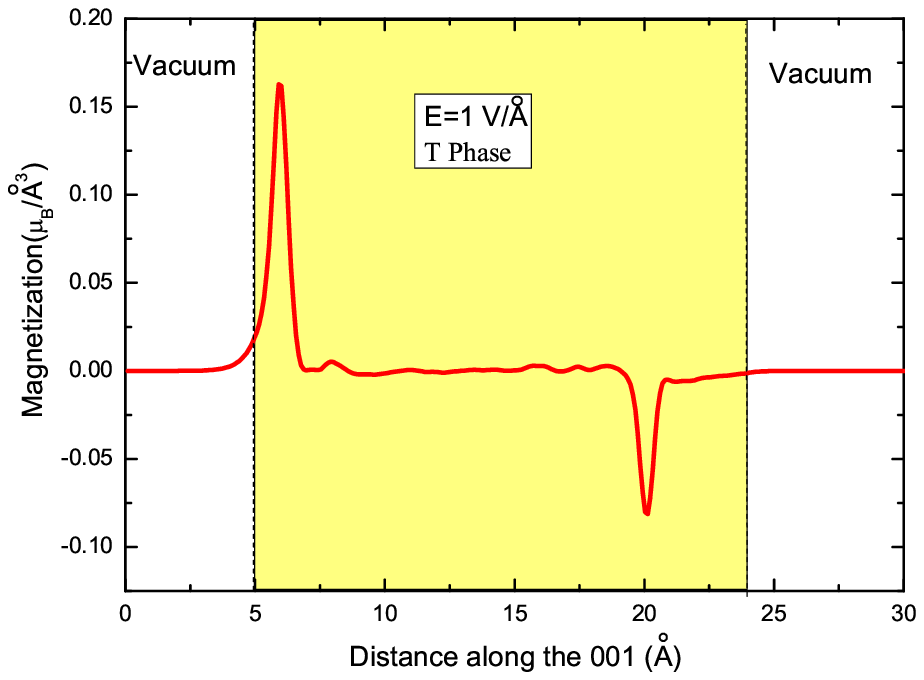}
\caption{ Distribution of magnetization across the T phase ultrathin BFO film averaged over the plane parallel to the film under electric field of E=1 V/${\AA}$.}
\end{figure}

\begin{figure}
\centering
\includegraphics[width=10 cm]{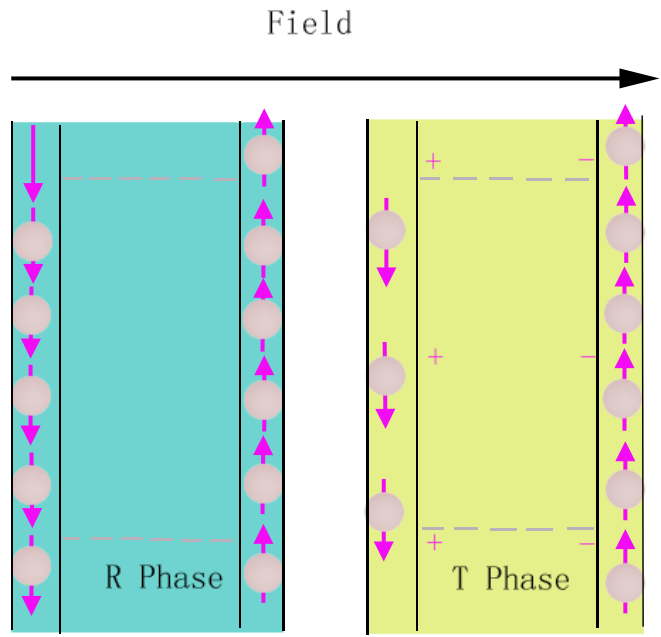}
\caption{ Spin transfer mechanism applying electric field for R and T phase BFO film.}
\end{figure}

\end{document}